\documentclass[twocolumn,showpacs,amsmath,amssymb]{revtex4}

\usepackage{graphicx}% Include figure files
\usepackage{dcolumn}% Align table columns on decimal point
\usepackage{bm}% bold math
\usepackage{amssymb}
\usepackage{amsmath}

\usepackage{epsf}
\usepackage{subfigure}
\usepackage{epstopdf}
\newcommand{\mean}[1]{\langle #1 \rangle}

\renewcommand{\i}{{\rm i}}
\newcommand{\e}{{\rm e}}
\begin{document}

\title{Overdamping by weakly coupled environments}

\author{Massimiliano Esposito}
%\email{mesposit@ulb.ac.be}
\affiliation{Center for Nonlinear Phenomena and Complex Systems,\\
Universit{\'e} Libre de Bruxelles, Code Postal 231, Campus Plaine, B-1050
Brussels, Belgium.}
\author{Fritz Haake}%
%\email{}
\affiliation{Fachbereich Physik, Universit{\"a}t Duisburg-Essen, 45117
Essen, Germany.\\}

\date{\today}

\begin{abstract}
A quantum system weakly interacting with a fast environment usually
undergoes a relaxation with complex frequencies whose imaginary
parts are damping rates quadratic in the coupling to the
environment, in accord with Fermi's ``Golden Rule''. We show for
various models (spin damped by harmonic-oscillator or random-matrix
baths, quantum diffusion, quantum Brownian motion) that upon
increasing the coupling up to a critical value still small enough to
allow for weak-coupling Markovian master equations, a new relaxation
regime can occur. In that regime, complex frequencies lose their
real parts such that the process becomes overdamped. Our results
call into question the standard belief that overdamping is
exclusively a strong coupling feature.
\end{abstract}

\pacs{05.30.-d; 03.65.Yz; 76.20.+q.}

\keywords{Quantum statistical mechanics, Relaxation, Overdamping,
Master equation.}

\maketitle
%%%%%%%%%%%%%%%%%%%%%%%%%%%%%%%%%%%%%%%%%%%%%%%%%%%%%%%%%%%%%%%%%%%%
\section{Introduction \label{intro}}

The dynamics of an isolated and finite quantum system consists of a
reversible superposition of oscillations with (real) Bohr frequencies
$\omega_S$. In order to understand the irreversible processes
occurring in finite quantum systems, such as relaxation to equilibrium
or decoherence, one needs to take into account the interaction between
the system and its environment. The weak-interaction limit together with the
Markovian approximation already allow a good understanding of such
irreversible processes and has some universal features. The generator
of the evolution of the (reduced) density matrix of the system
obtained by second-order perturbation theory (often called the
Redfieldian) is not an anti-Hermitian generator any more. Its
eigenvalues $\Gamma + {\i} \Omega$ acquire a real part $\Gamma$
describing irreversible decay to equilibrium. The imaginary parts of
the eigenvalues are shifted Bohr frequencies $\Omega=\omega_S-\delta
\omega$. The two shifts $\Gamma$ and $\delta \omega$, normally
increase (quadratically) as the strength of the coupling grows.

We here propose to show that Markovian perturbative master equations
such as the Redfield equation
\cite{Red,Kampen,HaaSpring,KuboB2,GaspRed,BreuPet,Esposito} allow for more
than just describing the well known normal damping just mentioned.
When the coupling strength is increased, it can happen at a critical
value that a shifted frequency $\Omega$ vanishes and for yet stronger
coupling goes imaginary. The pertinent eigenvalues $\Gamma - \vert
\Omega \vert$ are real and, interestingly, decrease with growing
coupling. The resulting relaxation is non-oscillatory, i.e.
overdamped. The principle purpose of this paper is to show that {\it
contrary to common belief} the transition to overdamping is
still compatible with perturbative treatment. In brief, {\it
overdamping can be a weak-coupling effect}.

All models to be studied here have Hamiltonians like
\begin{eqnarray}
\hat{H}&=&\hat{H}_{S}+\hat{H}_{B}+ \hat{S} \hat{B}\, ,
\label{A0000}
\end{eqnarray}
where $\hat{H}_{S}$ and $\hat{H}_{B}$ respectively generate the free
motion of the system and the environment (bath) while the interaction
involves respective coupling agents $\hat{S}$ and $\hat{B}$.

It may be well to emphasize that the so-called rotating-wave
approximation \cite{Gardiner,BreuPet}, extremely useful as it may be
for very weak damping, in particular in quantum optics, is definitely
not allowable for strong damping and overdamping. Indeed, the
rotating-wave approximation is based on the assumption that the Bohr
frequencies of the system are very large compared to the system
damping rate such that all ``anti-resonant'' terms can be time
averaged out when writing the master equation in the interaction
picture. But overdamping occurs precisely when the Bohr frequencies
of the system become of the order of or smaller then the system
damping rate. In a recent study of low-quality resonators
\cite{Hacken}, the rotating-wave approximation was shown to be still
affordable for overlapping resonances. But the Hamiltonians to be
employed in the present paper must retain the ``anti-resonant'' terms
that the rotating-wave approximation would suppress.

A word on physical contexts where overdamping shows up is in order.
One such is diffusion, a topic to be dealt with below (Section
\ref{diffusion}). Another one is temporal fluctuations in critical
phenomena, described by time dependent Ginzburg-Landau equations
without inertial terms \cite{HHM}; Ref.~\cite{Bausch} describes a
derivation of such a Ginzburg-Landau equation from an underlying
unitary evolution of a ``larger'' system.

The plan of the paper is as follows: In section
\ref{spin-environment}, we solve the Redfield master equation for a
two-level system interacting with a general environment. When the
environment is made of harmonic oscillators (spin-boson model), we
show in subsection \ref{spin-boson} that the transition from normal
damping to overdamping occurs at a critical value of the coupling
which can be made arbitrarily small and therefore accessible to
perturbation theory. For environment operators $\hat{H}_{B}$ and
$\hat{B}$ modeled by random matrices from the so-called Gaussian
orthogonal ensemble (spin-GORM model), we show in subsection
\ref{spin-GORM} that weak-coupling overdamping is compatible with the
exact dynamics computed numerically. In section \ref{diffusion}, we
show that the transition from a non-diffusive to a diffusive regime,
recently identified for a particle traveling in a spatially extended
system while interacting with an environment, corresponds in fact to a
transition from normal damping to overdamping; that transition will
turn out amenable to perturbative analysis. Finally, in section
\ref{QBM} we study the transition from normal damping to overdamping
for a central harmonic oscillator interacting with a large collection
of harmonic oscillators (quantum Brownian motion). We show that
overdamping again allows for perturbative treatment, by comparison
with the exact results known for this model. Conclusion are drawn in
section \ref{conclusion}.

%%%%%%%%%%%%%%%%%%%%%%%%%%%%%%%%%%%%%%%%%%%%%%%%%%%%%%%%%%%%%%%%
\section{Damped spin \label{spin-environment}}
\subsection{Hamiltonian and Markovian master equation}
Any two-level system has the Pauli matrices
$\hat{\sigma}_x,\hat{\sigma}_y,\hat{\sigma}_z$ (together with unity)
as a complete set of observables. If such a ``spin'' interacts with a
general environment we may choose the Hamiltonian as
\begin{eqnarray}
\hat{H} = \frac{\hbar\omega_0}{2} \hat{\sigma}_{z} + \hat{H}_{B}
+ \hat{\sigma}_x \hat{B}\,.
\label{F0111}
\end{eqnarray}
Inasmuch as the interaction $\hat{\sigma}_x \hat{B}$ does not commute
with the Hamiltonians for the uncoupled spin and bath, it allows for
transitions between the unperturbed energy levels. Denoting the
means of the spin observables by
\begin{eqnarray}
x(t)=\textrm{Tr} \hat{\rho}(t) \hat{\sigma}_{x} ,\;\;
y(t)=\textrm{Tr} \hat{\rho}(t) \hat{\sigma}_{y},\;\;
z(t)=\textrm{Tr} \hat{\rho}(t) \hat{\sigma}_{y} \label{F0211}
\end{eqnarray}
we write the Redfield equation as
\cite{GaspRed,Esposito}
\begin{eqnarray}
\dot{z}(t) &=& 2 \Gamma\, (z(\infty) - z(t)) \label{F0311}\\
\dot{x}(t) &=& -\omega_0 y(t) \nonumber \\
\dot{y}(t) &=& \frac{(\Omega^2+\Gamma^2)}{\omega_0} x(t) - 2 \Gamma y(t)
\nonumber\,,
\end{eqnarray}
with the time dependent damping rate $\Gamma(t)$ and frequency
$\Omega(t)$ and the stationary inversion $z(\infty)$
\begin{eqnarray}
\Gamma(t)&=& \frac{2}{\hbar^2} \int_{0}^{t} d\tau \cos (\omega_0 \tau)
\; C(\tau) \label{F0511}\\
\Omega(t)^2+\Gamma(t)^2&=& \omega_0^2 +
\frac{4}{\hbar^2} \omega_0 \int_{0}^{t} d\tau \sin (\omega_0 \tau)
\; C(\tau) \nonumber \\
\Gamma(t) \; z(\infty) &=&
\frac{2}{\hbar^2} \int_{0}^{t} d\tau \sin (\omega_0 \tau) \; D(\tau)
\nonumber\, .
\end{eqnarray}
Properties of the bath are represented by the functions $C(t)$ and
$D(t)$, respectively the real and imaginary parts of the equilibrium
autocorrelation function $\alpha(t)=\langle B(t)B(0)\rangle$ of the
bath coupling agent $B$ (For definition and properties see appendix
\ref{harmoniccorrel}).

The Markovian approximation consists in taking the upper bounds of the
time integrals in (\ref{F0511}) to infinity, such that the damping
constant and frequency become time independent, $\Gamma(\infty)\equiv
\Gamma,\Omega(\infty)\equiv \Omega\,$. That approximation is
legitimate when the spin dynamics characterized by the rates
$\omega_0,\Omega,\Gamma$ is much slower than the decay of the bath correlation
function $\alpha(t)$ and requires that we restrict the further
discussion to times much larger than the bath correlation time. We
may then rewrite (\ref{F0511}) as
\begin{eqnarray}
\Gamma&=&
\frac{\pi}{\hbar^2} (\tilde{\alpha}(\omega_0)+\tilde{\alpha}(-\omega_0))
\label{F0411}\\
\Omega^2+\Gamma^2&=& \omega_0^2 + \frac{4}{\hbar^2} \omega_0^2 \int d\omega
{\cal P} \frac{\tilde{\alpha}(\omega)}{\omega_0^2-\omega^2} \nonumber \\
z(\infty)&=& \frac{\tilde{\alpha}(-\omega_0)-\tilde{\alpha}(\omega_0)}
{\tilde{\alpha}(-\omega_0)+\tilde{\alpha}(\omega_0)}\, ,\nonumber
\end{eqnarray}
with $\tilde{\alpha}(\omega)$ the Fourier transform of $\alpha(t)$.
The solutions of equations (\ref{F0311}) in the Markovian limit read
\begin{eqnarray}
z(t)&=& z(\infty) + (z(0) - z(\infty)) \,\e^{-2 \Gamma t} \label{F0611}\\
x(t)&=&\frac{x(0) \Gamma - y(0) \omega_0}{\Omega}
\sin \left( \Omega t \right) \e^{-\Gamma t} \nonumber \\
&&+x(0) \cos \left( \Omega t \right) \e^{-\Gamma t} \nonumber \\
y(t)&=&\frac{x(0) ((\Omega^2+\Gamma^2)/\omega_0) - y(0) \Gamma}{\Omega}
\sin \left( \Omega t \right)
\e^{-\Gamma t} \nonumber \\
&&+ y(0) \cos \left( \Omega t \right) \e^{- \Gamma t} . \nonumber
\end{eqnarray}
The reduced density matrix $\rho=\frac{1}{2}+x\hat{\sigma}_x+y\hat{\sigma}_y+
z\hat{\sigma}_z$ can thus be written as a
superposition of four modes,
\begin{eqnarray}
\hat{\rho}(t)=\sum_{\xi=1}^{4} c_{\xi}(0) \,\hat{\rho}^{\xi}\,
\e^{s_{\xi} t}\,.
\label{F0612}
\end{eqnarray}
For normal damping, $s_1=0$, $s_2=-2 \Gamma$, $s_3=-\Gamma+{\i}\Omega$
and $s_4=-\Gamma-{\i}\Omega$.
Overdamping occurs when
\begin{eqnarray}
\Omega^2 < 0\,, \label{F0711}
\end{eqnarray}
and then the rates of (\ref{F0612}) are given by $s_1=0$,
$s_2=-2 \Gamma$, $s_3=-\Gamma+ \vert \Omega \vert$
and $s_4=-\Gamma- \vert \Omega \vert$.

%%%%%%%%%%%%%%%%%%%%%%%%%%%%%%%%%%%%%%%%%%%%%%%%%%%%%%%%
\subsection{The spin-boson model}\label{spin-boson}

Taking the bath as a collection of harmonic oscillators
\cite{Leggett,GaspRed} we have for its free Hamiltonian and
coupling agent
\begin{eqnarray}
\hat{H}_B &=&
\frac{1}{2} \sum_{n=1}^{N} (\hat{P}^2_n +\omega_n^2 \hat{Q}^2_n)
\,,\quad B=\sum_{n=1}^{N} \epsilon_n \hat{Q}_n . \label{F0811}
\end{eqnarray}
We assume a quasi-continuum of bath frequencies $\omega_n$,
employ a spectral function
$\gamma(\omega)=\sum_n\epsilon_n^2\delta(\omega_n-\omega)$, and adopt
Ullersma's choice [see \cite{Ullersma} and Appendix
\ref{harmoniccorrel}],
\begin{equation}\label{Ullspecstrength}
\gamma(\omega)=\frac{2}{\pi}\frac{\kappa\alpha^2\omega^2}{\alpha^2+\omega^2}
\,,
\end{equation}
where $\alpha$ is the decay rate of the autocorrelator
of the bath coupling agent and $\kappa$ an overall coupling strength.
Thus equipped we can evaluate the rates in (\ref{F0511}). In
the limits of high temperature, i.e.
$\beta\hbar\omega_0\equiv \hbar\omega_0/k_BT \ll 1$, we get
\begin{eqnarray}
\Gamma&\stackrel{\beta \to 0}{=}&
2 \frac{\kappa \alpha^2}{\beta \hbar^2 (\alpha^2+\omega_0^2)}\,
\stackrel{\frac{\omega_0}{\alpha} \to 0}{=} \,
\frac{2 \kappa}{\beta \hbar^2}
\label{F0911}\\
\Omega^2+\Gamma^2&\stackrel{\beta \to 0}{=}&
\omega_0^2+4 \frac{\kappa \alpha \omega_0^2}
{\beta \hbar^2 (\alpha^2+\omega_0^2)}
\stackrel{\frac{\omega_0}{\alpha}\to 0}{=} \omega_0^2
\label{F0913} \\
\Gamma \; z(\infty) &=& - \frac{\kappa \alpha^2 \omega_0}
{\hbar (\alpha^2+\omega_0^2)}
\stackrel{\frac{\omega_0}{\alpha} \to 0}{=} - \frac{\kappa \omega_0}{\hbar}
\,;\label{F0915}
\end{eqnarray}
here the limit $\omega_0 /\alpha \to 0$ has been taken to remain consistent
with the Markovian approximation; note that in the present section
$\kappa$ has the dimension of an action, such that $\Gamma$ is a rate.

The critical value of $\kappa$ at which overdamping occurs is now
found with the help of Eq. (\ref{F0711}) by subtracting Eq.(\ref{F0911})
to the power two to Eq. (\ref{F0913}). We find
\begin{eqnarray}
\kappa_c&\stackrel{\beta\to 0}{=}&\frac{\hbar^2 \beta \omega_0}{2}
\label{F1111}
\end{eqnarray}
If $\kappa > \kappa_c$, and if $\kappa_c$ is small enough to be
treated by perturbation theory we have a selfconsistent theory of
overdamping. Clearly, high temperatures are favorable for
that theory to apply since the pertinent $\kappa_c$ is suppressed by
the factor $\beta \hbar\omega_0 \ll 1$.

One might fear that our way of obtaining $\kappa_c$ is not completely
consistent if solely restricted to second-order perturbation theory
because $\Gamma^2$ is of order $\kappa^2$ while $\Omega^2+\Gamma^2$
is of order $\kappa$ and does not include the $\kappa^2$ corrections.
That fear would be eased by the following argument.
If we were to add ${\cal O}(\kappa^2)$ corrections to the right-hand sides
of Eqs. (\ref{F0911},\ref{F0913}), the results
(\ref{F1111}) for the critical coupling would be generalized
to series in powers of the leading terms displayed in (\ref{F1111}).
(In fact, Jang et al. Ref.\cite{Silbey} found $\Omega^2+\Gamma^2=
\omega_0^2(1+4\kappa^2/\hbar^2)$; by recalculating $\kappa_c$, we
again find Eq. (\ref{F1111}) if $\beta \hbar \omega_0 \ll 1$.)

Another look at the high-temperature rates reveals an interesting
feature of overdamping. We have from (\ref{F0612})
\iffalse
In the normal damping regime, the
rates given by (\ref{F0612}) become
\begin{eqnarray}
s_1&=&0 \label{F1113} \\
s_2&=&- \frac{4 \kappa}{\hbar^2 \beta} \nonumber \\
s_3&=&- \frac{2 \kappa}{\hbar^2 \beta} + i \frac{2 \kappa}{\hbar^2 \beta}
\sqrt{(\omega_0 \frac{\hbar^2 \beta}{2 \kappa})^2-1} \nonumber \\
s_4&=&- \frac{2 \kappa}{\hbar^2 \beta} - i \frac{2 \kappa}{\hbar^2 \beta}
\sqrt{(\omega_0 \frac{\hbar^2 \beta}{2 \kappa})^2-1} \nonumber .
\end{eqnarray}
Whether in the overdamping regime, the decay rates of the
subsystem become
\fi
\begin{eqnarray}
s_1&=&0 \label{F1114} \\
s_2&=&- \frac{4 \kappa}{\hbar^2 \beta} \nonumber \\
s_3&=&- \frac{2 \kappa}{\hbar^2 \beta} + \frac{2 \kappa}{\hbar^2 \beta}
\sqrt{1-\left(\frac{\kappa_c}{\kappa}\right)^2} \nonumber \\
%&=&- \frac{\beta \Delta^2}{4 \kappa}
%+ O(\frac{\hbar^2 \beta^3 \Delta^4}{\kappa^3}) \nonumber \\
s_4&=&- \frac{2 \kappa}{\hbar^2 \beta} - \frac{2 \kappa}{\hbar^2 \beta}
\sqrt{1-\left(\frac{\kappa_c}{\kappa}\right)^2} \nonumber \,.
%&=&-\frac{4 \kappa}{\hbar^2 \beta} + \frac{\beta \Delta^2}{4 \kappa}
%+ O(\frac{\hbar^2 \beta^3 \Delta^4}{\kappa^3}) \nonumber .
\end{eqnarray}
Most remarkably, the slowest relaxation rate of the spin, $\vert {\rm
Re}[s_3] \vert$, decreases when the coupling to the environment increases.
For strong overdamping, $\kappa_c/\kappa\ll 1$, we even have
\begin{eqnarray}
s_3 = -\frac{\beta \omega_0^2}{4 \kappa}
+ {\cal O}(\frac{\hbar^2 \beta^3 \omega_0^4}{\kappa^3}) \; .
\end{eqnarray}
This is in contrast to the normal-damping case, accessible from the
above by replacing $\sqrt{-1}\to +{\i}$, where the two slowest rates
$\vert {\rm Re}[s_2] \vert$ and $\vert {\rm Re}[s_3] \vert$ increase
as the coupling becomes stronger.

%%%%%%%%%%%%%%%%%%%%%%%%%%%%%%%%%%%%%%%%%%%%%%%%%%%%%%%%
\subsection{The spin-GORM model}\label{spin-GORM}

We retain the overall Hamiltonian (\ref{F0111}) but modify the
environment so as to let the free-bath Hamiltonian $\hat{H}_B$ and the
coupling agent $\hat{B}$ be represented by random matrices from the
Gaussian orthogonal ensemble (GOE). The resulting spin-GORM model was
studied in Refs.~\cite{Esposito,EspoGasp2}. We use the results of that
work; in particular, we adopt a unit of time that makes the
Hamiltonian dimensionless and the bath correlation time of order
$\hbar$ (see Eq.(\ref{G0100}) below). Specifically, we write
\begin{eqnarray}
\hat{H}_B = \frac{\hat{X}}{\sqrt{8N}} \,,\quad
\hat{B} = \eta \frac{\hat{X}^{'}}{\sqrt{8N}}\,;
\label{G0000}
\end{eqnarray}
here $\hat{X}$ and $\hat{X}^{'}$ are random $\frac{N}{2}
\times \frac{N}{2}$ GOE matrices with mean zero. Their
non-diagonal (resp. diagonal) elements have standard deviation
$\sigma_{ND}= 1$ (resp. $\sigma_{D} = \sqrt{2}$).
The parameter $\eta$ serves as a coupling strength.

To study this model it is convenient to assume that the environment is
initially in a microcanonical distribution with the (dimensionless)
energy $\epsilon$. The autocorrelator of the bath coupling agent then
reads
\begin{eqnarray}
\alpha(\epsilon,t)
\stackrel{N \to \infty}{=} \eta^2 \frac{J_1(t/(2\hbar))}{4 t/\hbar}\,
\e^{{\i} \epsilon t/\hbar} \label{G0100}
\end{eqnarray}
and has the Fourier transform
\begin{eqnarray}
\tilde{\alpha}(\epsilon,\omega)
\stackrel{N \to \infty}{=} \frac{\eta^2 \hbar}{2 \pi}
\sqrt{\frac{1}{4}-(\epsilon+ \hbar \omega)^2}\,.
\label{G0200}
\end{eqnarray}
It may be well to note that we here meet Wigner's semi-circle law
for the mean level density of the GOE.

\begin{figure}[h]
\centering
\rotatebox{0}{\scalebox{0.7}{\includegraphics{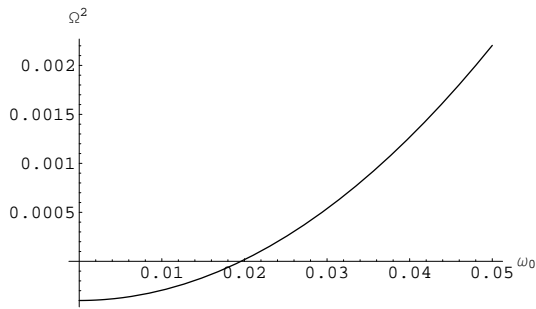}}}
\caption{$\eta=0.2$, $\epsilon=0$ and $\hbar=1$. This figure shows that the
condition for overdamping can be satisfied at weak coupling if $\omega_0$
is sufficiently small. This is still true for any generic choice of
$\epsilon$.} \label{fig1}
\end{figure}

The general rates of the Markovian Redfield equation given in Eq.
(\ref{F0411}) can be evaluated and read
\begin{eqnarray}
\Gamma(\epsilon)
= \frac{\eta^2}{2 \hbar} \!\left[ \!\sqrt{\frac{1}{4}-(\epsilon
-\hbar \omega_0)^2} +
\sqrt{\frac{1}{4}-(\epsilon+\hbar \omega_0)^2} \right]
\label{G0300}
\end{eqnarray}
and
\begin{eqnarray}
\Omega(\epsilon)^2+\Gamma(\epsilon)^2&=& \omega_0^2
+ \eta^2 \omega_0^2 \label{G0400} \\
&&\hspace{-2.5cm}- \frac{\eta^2}{\hbar} \omega_0
\frac{\sqrt{(\epsilon+\hbar \omega_0)^2-\frac{1}{4}}}{\pi}
\arctan \left( \frac{(\epsilon+\hbar \omega_0)+\frac{1}{2}}
{\sqrt{(\epsilon+\hbar \omega_0)^2-\frac{1}{4}}} \right) \nonumber \\
&&\hspace{-2.5cm}- \frac{\eta^2}{\hbar} \omega_0
\frac{\sqrt{(\epsilon+\hbar \omega_0)^2-\frac{1}{4}}}{\pi}
\arctan \left( \frac{(\epsilon+\hbar \omega_0)-\frac{1}{2}}
{\sqrt{(\epsilon+\hbar \omega_0)^2-\frac{1}{4}}} \right) \nonumber \\
&&\hspace{-2.5cm}+ \frac{\eta^2}{\hbar} \omega_0
\frac{\sqrt{(\epsilon-\hbar \omega_0)^2-\frac{1}{4}}}{\pi}
\arctan \left( \frac{(\epsilon-\hbar \omega_0)+\frac{1}{2}}
{\sqrt{(\epsilon-\hbar \omega_0)^2-\frac{1}{4}}} \right) \nonumber \\
&&\hspace{-2.5cm}+ \frac{\eta^2}{\hbar} \omega_0
\frac{\sqrt{(\epsilon-\hbar \omega_0)^2-\frac{1}{4}}}{\pi}
\arctan \left( \frac{(\epsilon-\hbar \omega_0)-\frac{1}{2}}
{\sqrt{(\epsilon-\hbar \omega_0)^2-\frac{1}{4}}} \right). \nonumber
\end{eqnarray}
If $\Omega^2< 0$, we have overdamping. To discuss that case, we
momentarily set $\Omega^2=A-B$ where $A=\Omega^2+\Gamma^2$ and is given by
(\ref{G0300}) and $B=\Gamma^2$ by (\ref{G0400}). When $\eta$ is large we
could have overdamping because $B>A$, but then perturbation theory may
fail and our approach lose selfconsistency. However, since all
terms in $A$ (but none in $B$) carry explicit factors $\omega_0$ or $\omega_0^2$,
and since the other quantities containing $\omega_0$ (i.e. $\sqrt{.}$ and
$\arctan(.)$) are bounded away from zero in the limit $\omega_0 \to 0$, it is always
possible to choose $\omega_0$ sufficiently small such that $A<B$ for small
$\eta$. This is illustrated in Fig. \ref{fig1}.

\begin{figure}[h]
\centering
\begin{tabular}{c}
\vspace*{0.5cm}
%\hspace*{-0.25cm}
\rotatebox{0}{\scalebox{0.7}{\includegraphics{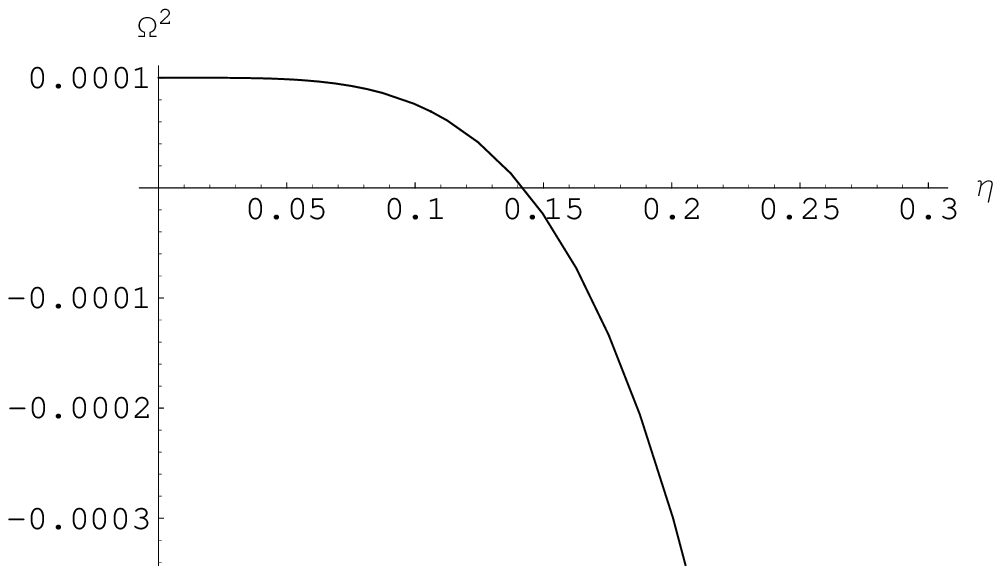}}} \hspace*{0.7cm}\\
\rotatebox{0}{\scalebox{0.7}{\includegraphics{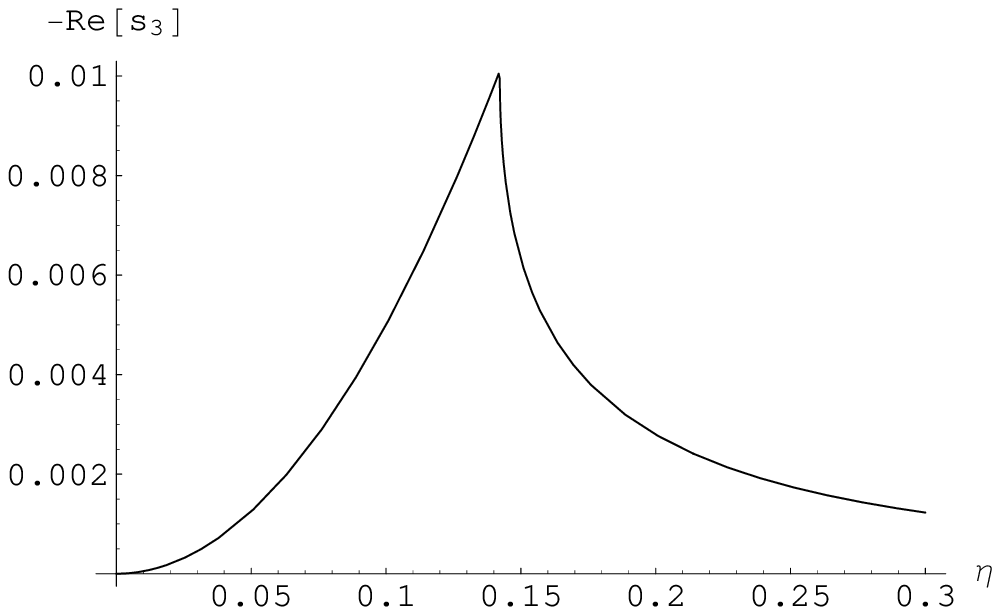}}}
\end{tabular}
\caption{$\omega_0=0.01$, $\epsilon=0$ and $\hbar=1$. The upper figure shows that
for a fixed and small value of $\omega_0$ their exist a critical and small
value of the coupling $\eta_c \approx 0.14$ above which overdamping occurs.
The lower figure illustrates the qualitative change of the coupling
dependence of the slowest relaxation rates when going from the
normal damping regime to the overdamped regime.}
\label{fig1b}
\end{figure}

The dependence of the smallest rates on the coupling strength is
similar as in the spin-boson model. The rates $|{\rm Re}[s_3]|$ and
$|{\rm Re}[s_4]|$ [see Eqs. (\ref{F0612})], grow with the coupling
constant $\eta$ in the normal-damping regime, have a cusp at the
transition, and then decay into the regime of overdamping, as
illustrated in Fig. \ref{fig1b}.

\begin{figure}[h]
\centering
\hspace*{-1.11cm}
\rotatebox{0}{\scalebox{0.55}{\includegraphics{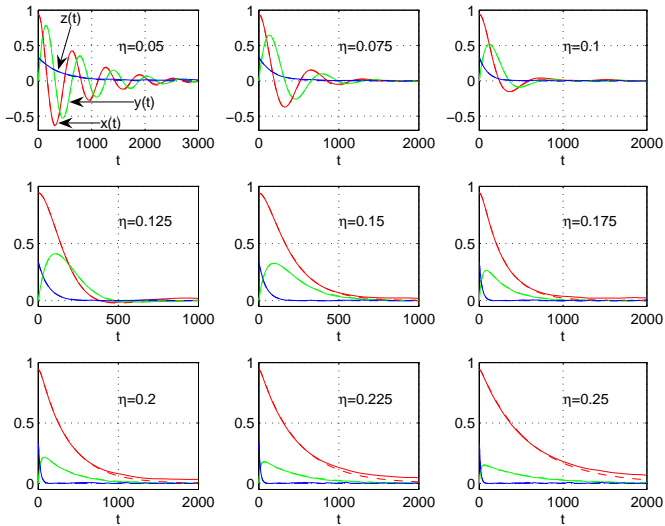}}}
\caption{(Color online) Transition from normal damping to overdamping in the
spin-GORM model.
The full lines represent the exact dynamics of the three spin observable
$x(t),y(t),z(t)$ obtained numerically by diagonalizing the full Hamiltonian
and the dashed lines represent the dynamics predicted by the Redfield equation
(second order perturbation theory).
The two results give curves which are so close to each other that the dashed
lines are almost invisible.
The situation depicted here is the same as in Fig. \ref{fig1b}
where $\eta$ varies and $\omega_0=0.01$, $\epsilon=0$ and $\hbar=1$.
As predicted by Redfield theory, the transition occurs at $\eta_c \approx 0.14$.
The initial condition is $x(0)=\sqrt{8}/3,y(0)=0,z(0)=1/3$.
For the exact dynamics we have taken $N=3000$ and a width of the
initial energy shell $\delta \epsilon=0.025$.}
\label{fig2}
\end{figure}

We have numerically solved the exact dynamics in order to verify that the
perturbative equation predicts the correct dynamics for normal damping as
well as for overdamping. The agreement is excellent as illustrated in
Fig. \ref{fig2}.
We can conclude that the spin-GORM model allows for overdamping at weak coupling.

%%%%%%%%%%%%%%%%%%%%%%%%%%%%%%%%%%%%%%%%%%%%%%%%%%%%%%%%%%%%%%%%%%%%%%%%%%%%%%%%%%%%%%%%%%%%%%%%%%%%%
\section{Diffusion model \label{diffusion}}

We now consider a particle moving on one dimensional closed
loop while interacting with an environment. The pertinent dynamics has
been studied recently in Refs.
\cite{Esposito,EspoGaspdiff1,EspoGaspdiff2} by using the Redfield
equation. A transition from nondiffusive to diffusive
relaxation has been identified. We shall here use the
results of this study to show that the transition mentioned in
fact is one from normal damping to overdamping.

The Hamiltonian of the loop constituting the subsystem is represented by an
$N\times N$ matrix
\begin{eqnarray}
\hat{H}_{S}=
\left(\begin{array}{ccccccc}
E_0 & -A & 0 & 0 & \hdots & 0 & -A \\
-A & E_0 & -A & 0 & \hdots & 0 & 0 \\
0 & -A & E_0 & -A & & 0 & 0 \\
\vdots & &\ddots &\ddots & \ddots & & \vdots \\
0 & 0 & & -A & E_0 & -A & 0 \\
0 & 0 & \hdots & 0 & -A & E_0 & -A \\
-A & 0 & \hdots & 0 & 0 & -A & E_0 \\
\end{array} \right)_{N \times N} \label{Diff0000}
\end{eqnarray}
taken in the site basis ${\vert l \rangle}$, where $l=0,1,\hdots,N-1$
labels the $N$ sites on the loop. The diagonal elements of $\hat{H}_S$
are the on-site energies of the particle while the
offdiagonal elements generate hopping to neighboring sites.

A weak interaction with an environment is described by the Redfield
master equation. The correlation time of the environment is assumed
much shorter than all characteristic time scales of the loop and
therefore the correlation function of the environment can be modeled
by
\begin{eqnarray}
\alpha_{l l'}(\tau)= 2\; Q \; \delta (\tau)\; \delta_{l l'}\, .
\label{Diff0100}
\end{eqnarray}
By using the Bloch theorem, the Redfield generator (containing $N^4$
elements) can be simplified in $N$ independent sectors (with $N^2$
elements), corresponding each to a given value of the Bloch number
$q$. For our finite loop periodicity yields $q = n 2 \pi / N$, where
$n=1,2,\hdots,N$. By diagonalizing a given sector we get $N$ eigenvalues
depending on $q$. The complete spectrum of the Redfield generator then
consists of the $N^2$ eigenvalues obtained by varying $q$.

As already mentioned, two relaxation regimes have been identified
in this model. In the nondiffusive regime all eigenvalues are complex
with real parts of similar magnitude, proportional to the coupling constant $Q$,
\begin{eqnarray}
{\rm Re}[s] \approx - \frac{2 Q}{\hbar^2} + {\cal O}(\frac{1}{N})\,.
\label{Diff0200}
\end{eqnarray}
However, in a given sector (therefore at a given $q$) when the coupling
term is increased beyond the value $Q = 2 \hbar A \sin \frac{q}{2}$,
one of the $N$ eigenvalues separates from the other $N-1$ ones.
This eigenvalue is always real and is called the diffusive one.
The diffusive branch is made of the diffusive eigenvalues
of the different sectors. These eigenvalues have a smaller magnitude
than the real parts of the nondiffusive eigenvalues.
They therefore control the long time relaxation of the subsystem.
The diffusive eigenvalues are given by
\begin{eqnarray}
s = - \frac{2 Q}{\hbar^2} + \frac{2 Q}{\hbar^2}
\sqrt{1- \left(\frac{2 \hbar A}{Q} \sin \frac{q}{2} \right)^2} \label{Diff0300}.
\end{eqnarray}
Lets define $Q_c \equiv 2 \hbar A \sin \frac{\pi}{N} \approx 2 \pi \hbar A / N$.
For $Q < Q_c$ no diffusive eigenvalues are present in the spectrum
and the relaxation regime is nondiffusive [see Eq. (\ref{Diff0200})].
As soon as $Q > Q_c$, at least two diffusive eigenvalues exist in the spectrum
and the relaxation regime is called the diffusive regime.
The two smallest diffusive eigenvalues controlling the long time scale
relaxation are
\begin{eqnarray}
s = - \frac{4 \pi^2 A^2}{Q N^2}\, .
\label{Diff0400}
\end{eqnarray}
Notice that the perturbative approach is consistent, because $Q_c$ can
be made as small as desired by choosing $A/N$ small.

It is already clear at this point that the nondiffusive (resp.
diffusive) regime implies normal damping (resp. overdamping). Indeed,
as for normal damping (resp. overdamping), the smallest relaxation
rates increase (resp. decrease) with growing coupling in the
nondiffusive (resp. diffusing) regime. Furthermore, as in the normal
damping (resp. overdamping) regime, the small Redfield eigenvalues are
complex (resp. real) in the nondiffusive (resp. diffusing) regime. We
can make that association even clearer if we assume the environment
made of harmonic oscillators which we model by using Ullersma's
spectral density [see appendix (\ref{harmoniccorrel})]. In this case,
we find that at high temperature, the zero-frequency limit of the
Fourier transform of the environment correlation function is given by
\begin{eqnarray}
\lim_{\omega \to 0} \tilde{\alpha}(\omega)
= \lim_{\omega \to 0} \frac{J(\omega)}{\omega}
= \frac{\kappa}{\pi \beta}\label{Diff0500}\,.
\end{eqnarray}
Since our instantaneous-decay assumption
(\ref{Diff0100}) implies
\begin{eqnarray}
\tilde{\alpha}(0)=\frac{Q}{\pi}\label{Diff0600}
\end{eqnarray}
we conclude
\begin{eqnarray}
Q=\frac{\kappa}{\beta} \label{Diff0700}
\end{eqnarray}
and thus find the diffusive-branch eigenvalues
\begin{eqnarray}
s &=& - \frac{2 \kappa}{\hbar^2 \beta} + \frac{2 \kappa}{\hbar^2 \beta}
\sqrt{1- (4 \hbar A \sin (\frac{q}{2}) \frac{\beta}{2 \kappa})^2}
\label{Diff0800}\\ &=& -\frac{\beta A^2}{\kappa} q^2
+{\cal O}(\frac{\hbar^2 \beta^3 A^4}{\kappa^3} q^4). \nonumber
\end{eqnarray}
The similarity between these diffusive eigenvalues and the smallest
eigenvalue of the spin-boson model in the overdamping regime [$s_3$ in
Eq. (\ref{F1114})] is obvious, as is similarity between the real part
of the small nondiffusive eigenvalues [(\ref{Diff0200}) with
(\ref{Diff0700})] and the real parts of the small eigenvalues
of the spin-boson model in the normal damping regime [$s_3$ 
and $s_4$ in (\ref{F1114})].

%%%%%%%%%%%%%%%%%%%%%%%%%%%%%%%%%%%%%%%%%%%%%%%%%%%%%%%%%%%%%%%
\section{Quantum Brownian Motion (QBM) \label{QBM}}
\subsection{Hamiltonian}
In this section we study overdamping in an exactly solvable model of
Brownian motion. The model is made of a central harmonic oscillator
interacting with an environment which itself is a collection of
harmonic oscillators (see, e.g. \cite{Ullersma,Haake,Zurek,BreuPet};
these references will lead the reader to earlier work). The exact solution
proves extremely valuable for our endeavor since it will be seen to yield,
in the Markovian limit, precicely the same condition for overdamping as the
perturbative treatment.

We write the total QBM Hamiltonian as \cite{Rem1}
\begin{eqnarray}
\hat{H}=\frac{1}{2} (\hat{P}^2 + \omega_0^2 \hat{Q}^2) +
\frac{1}{2} \!\sum_{n=1}^{N}\! \!\Big(\!\hat{P}^2_n +\omega_n^2 (\hat{Q}_n
-\frac{\epsilon_n}{\omega_n^2}\hat{Q})^2\!\Big)
\label{A0100}
\end{eqnarray}
and thus have the system and bath parts
\begin{eqnarray}
\hat{H}_{S}&=&\frac{1}{2} \Big(\hat{P}^2 +\big(\omega_0^2 + \sum_{n=1}^{N}
\frac{\epsilon_n^2}{\omega_n^2}\big) \hat{Q}^2\Big)\,,
\label{A0101}\\
%\end{eqnarray}
%The environment Hamiltonian is given by
%\begin{eqnarray}
\hat{H}_{B}&=&\frac{1}{2} \sum_{n=1}^{N}
\Big(\hat{P}^2_n +\omega_n^2 \hat{Q}^2_n\Big)\,.
\label{A0102}
\end{eqnarray}
The coupling agents of system and bath read
\begin{eqnarray}
\hat{S}= \hat{Q}_0 , \ \ ; \ \
\hat{B}= - \sum_{n=1}^{N} \epsilon_n \hat{Q}_n\,.
\label{A0103}
\end{eqnarray}

The QBM Hamiltonian (\ref{A0100}) is a sum of squares and thus
manifestly positive. A not manifestly positive variant of that
Hamiltonian \cite{Ullersma,Haake}, discussed in Appendix
\ref{oscillatorHR}, can be mapped onto the QBM
Hamiltonian by a renormalization of the bare frequency of the central
oscillator. That observation allows us to use the exact results of
Ref.~\cite{Haake} for our present study of QBM.

\subsection{Exact treatment}

The Hamiltonian (\ref{A0100}) generates the Heisenberg equations of motion
\begin{eqnarray}
\dot{\hat{P}}(t) &=&
- \Big(\omega_0^2+\sum_{n=1}^{N} \frac{\epsilon^2}{\omega_n^2}\Big)
\hat{Q}(t) - \sum_{n=1}^{N} \epsilon_n \hat{Q}_n(t)
\nonumber \\
\dot{\hat{P}}_n(t) &=&
- \omega_n^2 \hat{Q}_n(t) - \epsilon_n \hat{Q}(t) \nonumber \\
\dot{\hat{Q}}(t) &=&
\hat{P}(t) \nonumber \\
\dot{\hat{Q}}_n(t) &=&
\hat{P}_n(t) \label{A0600}\,.
\end{eqnarray}
The solution of (\ref{A0600}) can be written as
\begin{eqnarray}
\hat{Q}_{\nu}(t)&=&\sum_{\nu=0}^{N} \Big( \dot{A}_{\mu\nu}(t)
\hat{Q}_{\nu}(0) +
A_{\mu\nu}(t) \hat{P}_{\nu}(0) \Big)\label{A0805} \\
\hat{P}_{\nu}(t)&=&\dot{\hat{Q}}_{\nu}(t) \nonumber \,;
\end{eqnarray}
the indices $\mu$ and $\nu$ step from $0$ to $N$ and
$\hat{Q}_0\equiv \hat{Q},\,\hat{P}_0\equiv\hat{P}$.
All $A_{\mu\nu}(t)$'s can be expressed in terms of the function
\begin{eqnarray}
g(z)&=&z^2-\omega_0^2-\sum_{n=1}^{N} \frac{\epsilon^2}{\omega_n^2}
-\sum_{n=1}^{N}\frac{\epsilon_n^2}{z^2-\omega_n^2}\, . \label{A0800}
\end{eqnarray}
The zeros of $g(z)$ yield the eigenfrequencies of Eqs. (\ref{A0600}).

Assuming the bath frequencies to form a quasi-continuum we employ a
spectral function
$\gamma(\omega)=\sum_n\epsilon_n^2\delta(\omega_n-\omega)$ to replace
the sum in (\ref{A0800}) by an integral,
\begin{eqnarray}
g(z)=z^2-\omega_0^2-\int_{0}^{\infty} d\omega \frac{\gamma(\omega)}
{\omega^2}-\int_{0}^{\infty} d\omega \frac{\gamma(\omega)}
{z^2-\omega^2}\,.
\label{A1100}
\end{eqnarray}
\iffalse
Ullersma's modeling of the spectral function (see
appendix \ref{harmoniccorrel}) entails
\begin{eqnarray}
g(z)&=&z^2-\omega_0^2-\kappa \alpha-\int_{0}^{\infty} d\omega
\frac{\gamma(\omega)} {z^2-\omega^2}\,.
\label{A1200}
\end{eqnarray}
\fi

We adopt an initial condition with statistical independence of central
oscillator and bath, without restriction for the density operator
$\rho(0)$ of the central oscillator,
\begin{eqnarray}
\hat{\rho}_{\rm tot}(0)
=\hat{\rho}(0) \frac{e^{-\beta \hat{H}_B}}{Z_B}\,.\label{A1502}
\end{eqnarray}
The time dependent density operator of
the central oscillator then obeys the exact master equation
\begin{eqnarray}
\dot{\hat{\rho}}(t) &=& -\frac{i}{2 \hbar} \lbrack \hat{P}^2 -
f_{pq}(t) \hat{Q}^2 , \hat{\rho}(t)
\rbrack \label{A1900} \\
&& + \frac{i}{\hbar} f_{pp}(t) \lbrack \hat{Q} , \lbrack \hat{P} ,
\hat{\rho}(t) \rbrack_+ \rbrack \nonumber \\
&& - \frac{1}{\hbar^2} d_{pp}(t) \lbrack \hat{Q} , \lbrack \hat{Q} ,
\hat{\rho}(t) \rbrack \rbrack \nonumber \\
&& + \frac{1}{\hbar^2} d_{pq}(t) \lbrack \hat{P} , \lbrack \hat{Q} ,
\hat{\rho}(t) \rbrack \rbrack \nonumber\,,
\end{eqnarray}
with $[\cdot,\cdot]_+$ the anticommutator. The drift and diffusion coefficients
$f_{pq}(t),f_{pp}(t),d_{pp}(t),d_{pq}(t)$ can be found in
\cite{Haake}; they can all be expressed in terms of
the quantity $A(t) \equiv A_{00}(t)$. To get an explicit result for that
amplitude we adopt Ullersma's spectral function,
\begin{equation}\label{Ullspecstrength}
\gamma(\omega)=\frac{2}{\pi}\frac{\kappa\alpha^2\omega^2}{\alpha^2+\omega^2}
\,,
\end{equation}
where $\alpha$ and $\kappa$ are the decay rate of the autocorrelator of the
bath coupling agent and an overall coupling strength, both now of the
dimension of a frequency. For that choice the amplitude in question
takes the form
\begin{eqnarray}
A(t)&=& \frac{2 \Gamma}{\lambda^2+\Omega^2+\Gamma^2- 2 \lambda
\Gamma} \left( e^{- \lambda t} - e^{- \Gamma t} \cos (\Omega t)
\right) \nonumber \\
&&+ \frac{\lambda^2+\Omega^2-\Gamma^2}{\lambda^2+\Omega^2+\Gamma^2
- 2 \lambda \Gamma} \frac{1}{\Omega} e^{- \Gamma t} \sin (\Omega t)\,.
\label{A1502}
\end{eqnarray}
Here, the three rates ($\Gamma$, $\Omega$, $\lambda$) control the
exact dynamics; they are connected to the three model parameters
($\omega_0$, $\kappa$, $\alpha$) by the characteristic equations
\begin{eqnarray}
&&\lambda = \alpha - 2 \Gamma , \nonumber \\
&&\omega_0^2 + \alpha \kappa = \Omega^2 + \Gamma^2 + 2 \lambda \Gamma\, ,
\label{A1700}\\
&&\omega_0^2 = (\Omega^2 + \Gamma^2) (\lambda/\alpha)\, .\nonumber
\end{eqnarray}
The coupling between central oscillator and bath is thus seen
to shift the unperturbed frequency as $\omega_0\to \Omega+{\i}\Gamma$
and the unperturbed bath decay rate as $\alpha\to\lambda$.

We should mention that the (diffusion) coefficients $d_{pp}(t)$ and
$d_{pq}(t)$, in contrast to (the drift coefficients) $f_{pq}(t)$ and
$f_{pp}(t)$, also depend on the temperature.

As a final comment on the exact solution of the model we would like to
add that, due to the initial condition (\ref{A1502}), we have
$\mean{\hat{P}_n}=\mean{\hat{Q}_n}=0$ and therefore get the mean
displacement of the central oscillator from
(\ref{A0805}) as
\begin{eqnarray}
\mean{\hat{Q}(t)}= \dot{A}(t) \mean{\hat{Q}(0)}
+ A(t) \mean{\hat{P}(0)}\,. \label{A1501}
\end{eqnarray}

Turning to the Markovian limit we assume that environment
correlations decay fast relative to the time scales of the
central oscillator. In technical terms, we require
\begin{equation}\label{Markov}
\alpha,\lambda \gg \vert \Gamma + {\i} \Omega \vert\,.
\end{equation}
That Markovian limit does not imply
weak coupling. The characteristic equations
(\ref{A1700}) now become
\begin{eqnarray}
&&\lambda = \alpha \nonumber \\
&&\omega_0^2 + \alpha \kappa = \Omega^2 + \Gamma^2 + 2 \alpha \Gamma
\label{A2100}\\
&&\omega_0^2 = (\Omega^2 + \Gamma^2) \nonumber
\end{eqnarray}
and entail the explicit results
\begin{eqnarray}
\Gamma = \frac{\kappa}{2}\,, \quad
\Omega^2 = \omega_0^2 - \frac{\kappa^2}{4}\,,\quad \lambda=\alpha
\label{A2301}\,.
\end{eqnarray}
The master equation now reads, for times
$t\gg\alpha^{-1}$,
\begin{eqnarray}
\dot{\hat{\rho}}(t) &=& -\frac{\i}{2 \hbar} \lbrack \hat{P}^2 + \omega_0^2
\hat{Q}^2 , \hat{\rho}(t) \rbrack \label{A1900} \\
&& - \frac{\i}{\hbar} \Gamma \lbrack \hat{Q} ,
\lbrack \hat{P} , \hat{\rho}(t)
\rbrack_+ \rbrack \nonumber \\
&& - \frac{2}{\hbar^2} \Gamma \langle \hat{P}^2\rangle_{\rm eq} \lbrack
\hat{Q} ,
\lbrack \hat{Q} , \hat{\rho}(t) \rbrack \rbrack \nonumber \\
&& + \frac{1}{\hbar^2} \big( \omega_0^2 \langle\hat{Q}^2\rangle_{\rm eq}
- \langle\hat{P}^2\rangle_{\rm eq}\big)
\lbrack \hat{P} , \lbrack \hat{Q} , \hat{\rho}(t) \rbrack \rbrack
\nonumber\,.
\end{eqnarray}
The exact expressions for the stationary second moments $\langle\hat{Q}^2
\rangle_{\rm eq}$ and $\langle \hat{P}^2 \rangle_{\rm eq}$ are lengthly and can be
found in \cite{Haake}; they are completely characterized by the three
rates ($\Gamma$, $\Omega$, $\lambda$)
and by the temperature.\\

In the Markovian limit under study, the amplitude $A(t)$ in
(\ref{A1502}) also simplifies to
\begin{eqnarray}
A(t)&=& \frac{1}{\Omega} e^{- \Gamma t} \sin (\Omega t)\,,\quad
t\gg 1/\alpha\,.
\label{A1901}
\end{eqnarray}
We can now see that overdamping arises when $\Omega$ becomes a pure
imaginary number or equivalently when $\omega_0 < \Gamma$. The
transition between normal damping and
overdamping occurs at $\Omega=0$, for the
critical coupling
\begin{eqnarray}
\kappa_c=2 \omega_0 \label{A2700}\,.
\end{eqnarray}
That critical coupling will have to be compared with the one obtained
perturbatively.

\subsection{Perturbative treatment}
In order to compare exact and perturbative results we now look at the
Redfield master equation for the QBM Hamiltonian \cite{Zurek}
\begin{eqnarray}
\dot{\hat{\rho}}(t) &=& -\frac{i}{2 \hbar} \lbrack \hat{P}^2 +
(\Omega_{p}^2+\Gamma_{p}^2) \hat{Q}^2 , \hat{\rho}(t) \rbrack
\label{C0000} \\
&& - \frac{i}{\hbar} \Gamma_{p} \lbrack \hat{Q}, \lbrack \hat{P} ,
\hat{\rho}(t)
\rbrack_+ \rbrack \nonumber \\
&& - \frac{2}{\hbar^2} \Gamma_{p} \langle\hat{P}^2 \rangle_{\rm eq} \lbrack \hat{Q} ,
\lbrack \hat{Q}, \hat{\rho}(t) \rbrack \rbrack \nonumber \\
&& + \frac{1}{\hbar^2} \!\left(\!\!(\Omega_{p}^2+\Gamma_{p}^2)
\langle\hat{Q}^2\rangle_{\rm eq} -
\langle\hat{P}^2\rangle_{\rm eq}\! \right)
\lbrack \hat{P} , \lbrack \hat{Q} , \hat{\rho}(t) \rbrack \rbrack \nonumber
\end{eqnarray}
where
\begin{eqnarray}
&&\Gamma_{p} = \frac{1}{\hbar} \int_{0}^{t} d t
\frac{\sin \omega_0 t}{\omega_0}
\; D(t)\,, \label{C0100} \\
&&(\Omega_{p}^2+\Gamma_{p}^2) = \omega_{0}^2 + \!\int_{0}^{\infty}\! d\omega
\frac{\gamma(\omega)}{\omega^2} + \frac{2}{\hbar} \int_{0}^{t}
d t \cos \omega_0 t \; D(t)\,, \nonumber \\
&&2 \Gamma_{p} \langle P^2\rangle_{\rm eq} = \int_{0}^{t} d t \cos
\omega_0 t \; C(t)\,, \nonumber \\
&&(\Omega_{p}^2+\Gamma_{p}^2) \langle Q^2 \rangle_{\rm eq}
- \langle P^2 \rangle_{\rm eq} =
\int_{0}^{t} d t \frac{\sin \omega_0 t}{\omega_0}
\; C(t) . \nonumber
\end{eqnarray}
The Markovian approximation consists in taking the upper bounds of the
time integrals of (\ref{C0100}) to infinity and is justified when the
free motion of the central oscillator (characterized by the frequency
$\omega_0$) is much slower than the characteristic decay rate $\alpha$
of the correlation function of the environment ($\omega_0/\alpha \to
0$). Again using Ullersma's spectral function (\ref{Ullspecstrength})
we get the foregoing rates as
\begin{eqnarray}
&&\Gamma_{p} = \frac{\kappa \alpha^2}{2 (\alpha^2+\omega_0^2)}
= \frac{\kappa}{2}+{\cal O}(\frac{\omega_0^2}{\alpha^2}) \label{C0200} \\
&&(\Omega_{p}^2+\Gamma_{p}^2) = \omega_0^2 + \kappa \alpha -
\frac{\kappa \alpha^3}{\alpha^2+\omega_0^2} = \omega_0^2 +
{\cal O}(\frac{\kappa \omega_0^2}{\alpha}) \nonumber \\
&&\langle P^2 \rangle_{\rm eq} = \frac{\hbar \omega_0}{2} {\rm coth}
\frac{\beta
\hbar \omega_0}{2} \stackrel{\beta \to 0}{=} \frac{1}{\beta}
\nonumber\\
&&(\Omega_{p}^2+\Gamma_{p}^2) \langle Q^2\rangle_{\rm eq}
- \langle P^2\rangle_{\rm eq}
\stackrel{\beta \to 0}{=} \nonumber \\
&&\frac{\kappa \alpha}{\beta (\alpha^2+\omega_0^2)}
=\frac{\kappa}{\beta \alpha} \left( 1
+{\cal O}(\frac{\omega_0^2}{\alpha^2})\right) \nonumber
\end{eqnarray}
To be consistent with the Markovian assumption, all terms of order
$\omega_0/\alpha$ or smaller should be disregarded.
%(For a discussion
%of this point see Ref. \cite{Silbey} where the perturbative treatment
%is extended to fourth order in the coupling.)
When using the lowest-order master equation (\ref{C0000}) we recover the
mean displacement $\langle Q(t)\rangle$ of the rigorous treatment; in fact,
we even get coinciding results for the non-perturbative and the
perturbative rates in the Markovian limit $\omega_0/\alpha\to 0$, i.e.
$\Gamma=\Gamma_{p}=\kappa/2$ and
$\Omega=\Omega_{p}=\omega_0^2-\kappa^2/4$. In particular, therefore,
the transition to overdamping occurs at the same critical value of the
coupling, given by Eq. (\ref{A2700}). We conclude that the overdamping
regime in the QBM model in the Markovian limit can be described by
second-order perturbation theory and therefore is a weak-coupling
overdamping. It is worth mentioning that this result was anticipated
by Cohen-Tannoudji in \cite{Tannoudji}.

We finally note that for strong overdamping the slowest decay rate of
the QBM model reads
\begin{eqnarray}
s = - \frac{\kappa}{2} + \frac{\kappa}{2} \sqrt{1- (\frac{\kappa_c}{\kappa}
)^2} = -\frac{\omega_0^2}{\kappa} + {\cal O}(\frac{\omega_0^4}{\kappa^3})
\label{Diff0300},
\end{eqnarray}
in obvious similarity to the corresponding limit for the other models
studied above [see (\ref{F1114}) and (\ref{Diff0800})].

%%%%%%%%%%%%%%%%%%%%%%%%%%%%%%%%%%%%%%%%%%%%%%%%%%%%%%%%%%%%%%%%%%%%%%%%%%%%%%%%%%%%%%%%%%%%%%%%%%%%%
\section{Conclusion \label{conclusion}}

For four different models, made of a system weakly interacting with
its environment, we have studied the transition from normal damping to
overdamping. Normal damping has slowest relaxation rates that
increase with growing coupling strength and is characterized by
exponentially damped oscillations. In the overdamped regime the
smallest relaxation rates decrease with growing coupling and the
dynamics displays non-oscillatory exponential decay. The critical
value of the coupling at which the transition from normal damping to
overdamping occurs can often be made sufficiently small (by tuning
model parameters) to be describable by weak-coupling master equations
such as the Redfield equation. One way to make the critical coupling
small is to decrease the bare frequencies of the system, but other
parameters like the temperature of the system size can also enter the
game. The compatibility of weak coupling and overdamping is counter to
intuitive and widely spread expectations.

%%%%%%%%%%%%%%%%%%%%%%%%%%%%%%%%%%%%%%%%%%%%%%%%%%%%%%%%%%%%%%%%%%%%%%%%%%%%%%%%%%%%%%%%%%%%%%%%%%%%%
%%%%%%%%%%%%%%%%%%%%%%%%%%%%%%%%%%%%%%%%%%%%%%%%%%%%%%%%%%%%%%%%%%%%%%%%%%%%%%%%%%%%%%%%%%%%%%%%%%%%%

\begin{acknowledgments}
M. E. thanks Professor P. Gaspard for support and encouragement in
this research. M. E. is supported by the ``Minist{\`e}re de la
Culture, de l'Enseignement Sup{\'e}rieur et de la Recherche du
Grand-Duch{\'e} de Luxembourg". F. H. thanks Pierre Gaspard for
hospitality at the Universit{\'e} Libre de
Bruxelles which made this work possible.
\end{acknowledgments}

%%%%%%%%%%%%%%%%%%%%%%%%%%%%%%%%%%%%%%%%%%%%%%%%%%%%%%%%%%%%%%%%%%%%%%%%%%%%%%%%%%%%%%%%%%%%%%%%%%%%%

\appendix

%%%%%%%%%%%%%%%%%%%%%%%%%%%%%%%%%%%%%%%%%%%%%%%%%%%%%%%%%%%%%%%
\section{Harmonic oscillator environments}\label{harmoniccorrel}

We briefly recall some properties of the equilibrium autocorrelator of
the environment coupling agent $B$,
\begin{eqnarray}
\alpha(t) &=& \langle \hat{B}(t)\hat{B}\rangle= C(t) + {\i} D(t)\label{B0000}\\
&=& {\rm Tr}_B \hat{\rho}^{\rm eq}_B \e^{-{\i} \hat{H}_B t/\hbar} \hat{B}
\e^{{\i} \hat{H}_B t/\hbar} \hat{B} \,,\nonumber\\
\hat{\rho}^{\rm eq}_B&=&\e^{-\beta \hat{H}_B}/Z_B\label{caneq}\,.
\end{eqnarray}
The real and imaginary parts of $\alpha(t)$ obey $C(t)=C(-t)$ and
$D(t)=-D(-t)$. Their Fourier tarnsforms (defined as
$\tilde{\alpha}(\omega) = \frac{1}{2\pi}\int_{-\infty}^{\infty} dt
\e^{{\i} \omega t} \alpha(t) = \tilde{C}(\omega) + {\i}
\tilde{D}(\omega)$) are related by the fluctuation-dissipation theorem
\begin{eqnarray}
\tilde{C}(\omega)= 2{\i} \frac{E_{\beta}(\omega)}{\hbar \omega}
\tilde{D}(\omega) , \label{B0100}
\end{eqnarray}
where
\begin{eqnarray}
E_{\beta}(\omega)=\frac{\hbar \omega}{2} \coth{\frac{\beta \hbar \omega}{2}}
\label{B0200}
\end{eqnarray}
is the thermal energy of an oscillation with frequency $\omega$.
As a consequence, we can write our
correlator as
\begin{eqnarray}
\alpha(t) = \int_{0}^{\infty} d\omega \hbar J(\omega)
\Big({\rm coth} \frac{\beta \hbar \omega}{2} \cos \omega t
-{\i} \sin \omega t\Big)\, ,
\label{B0300}
\end{eqnarray}
thus introducing the spectral strength $J(\omega)$ of the environment
often used in the literature,
\begin{equation}
J(\omega)=\frac{2{\i}}{\hbar}\tilde{D}(\omega)\,,\quad \omega>0\,.
\end{equation}
It is in fact customary to use that spectral strength only for
positive frequencies; an extension to real frequencies could be to
require $J$ to be odd in $\omega$.

For an oscillator bath with
\begin{eqnarray}
\hat{H}_{B}=\frac{1}{2} \sum_{n=1}^{N}
(\hat{P}^2_n +\omega_n^2 \hat{Q}^2_n)\,,\quad
\hat{B}= \sum_{n=1}^{N} \epsilon_n \hat{Q}_n
\nonumber
\end{eqnarray}
the correlator becomes
\begin{eqnarray}
\alpha(t) &=& \sum_{n=1}^{N} \epsilon_n^2 {\rm Tr}_B
\frac{\e^{- \beta \hat{H}_B}}{Z_B} \hat{Q}_n(t) \hat{Q}_n(0)\nonumber \\
&=&\sum_{n=1}^{N} \frac{\hbar \epsilon_n^2}{2 \omega_n}
\Big({\rm coth} \frac{\beta \hbar \omega_n}{2} \cos \omega_n t
-{\i} \sin \omega_n t\Big)
\label{B0400}\\
&=& \int_{0}^{\infty} d\omega \frac{\gamma(\omega) \hbar}{2 \omega}
\Big({\rm coth} \frac{\beta \hbar \omega}{2} \cos \omega t
-{\i} \sin \omega t\Big)\nonumber
%\label{B0500}
\end{eqnarray}
and has the Fourier transform
\begin{eqnarray}
\tilde{\alpha}(\omega) = \frac{\gamma(\vert \omega \vert ) \hbar}
{4 \omega } ({\rm coth} \frac{\beta \hbar \omega}{2} + 1)\, .
\label{B0501}
\end{eqnarray}
A comparison of the general form (\ref{B0300}) with the
oscillator-bath form (\ref{B0400}) of the correlator $\alpha(t)$ shows that
the two spectral strengths $J(\omega)$ and $\gamma(\omega)$ (which are both 
common currency) are related as
\begin{eqnarray}
J(\omega) = \frac{\gamma(\omega)}{2 \omega}\,,\quad \omega>0\,. \label{B0600}
\end{eqnarray}

Ullersma's choice
\cite{Ullersma} (also called Drude strength)
\begin{eqnarray}
\gamma(\omega) &=& \frac{2}{\pi} \frac{\kappa \alpha^2 \omega^2}
{\alpha^2+\omega^2}\,.
\label{B0601}
\end{eqnarray}
corresponds to an ohmic environment because
at small frequencies $J(\omega) \sim \kappa \omega / \pi$.\\

At high temperatures, the real part of the environment correlator
is given by
\begin{eqnarray}
C(t) &\stackrel{\beta \to 0}{=}& \int_{0}^{\infty} d\omega
\frac{\gamma(\omega)}{\beta \omega^2} \cos \omega t
%&\stackrel{\beta \alpha \to 0}{=}&
=\frac{\kappa \alpha}{\beta} e^{-\alpha |t|}\,.
\label{B0700}
\end{eqnarray}
%while at low temperatures we have
%\begin{eqnarray}
%C(t) &\stackrel{\beta \to \infty}{=}& \int_{0}^{\infty} d\omega
%\frac{\hbar\gamma(\omega)}{2 \omega} \cos \omega t\,.
%\label{B0701}
%\end{eqnarray}
The imaginary part of the environment correlation function is
independent of temperature and reads
\begin{eqnarray}
D(t) &=& - \int_{0}^{\infty} d\omega \frac{\gamma(\omega)}
{2 \omega} \sin \omega t %\label{B0800} \\
= - \frac{\hbar \kappa \alpha^2}{2} e^{- \alpha |t|}\,{\rm sgn}(t)\,.
\nonumber
\end{eqnarray}

\iffalse
We finally list some integrals used in the paper
\begin{eqnarray}
\int_{0}^{\infty} d \tau \cos (\omega \tau) C(\tau)
&\stackrel{\beta \alpha \to 0}{=}& \frac{\kappa \alpha^2}
{\beta (\alpha^2+\omega^2)} \label{B0900} \\
\int_{0}^{\infty} d \tau \sin (\omega \tau) C(\tau)
&\stackrel{\beta \alpha \to 0}{=}& \frac{\kappa \alpha \omega}
{\beta (\alpha^2+\omega^2)} \label{B0901} \\
\int_{0}^{\infty} d \tau \cos (\omega \tau) C(\tau)
&\stackrel{\beta \to \infty}{=}& \frac{\hbar \kappa \alpha^2 \omega}
{(\alpha^2+\omega^2)} \label{B0902} \\
\int_{0}^{\infty} d \tau \sin (\omega \tau) C(\tau)
&\stackrel{\beta \to \infty}{=}& \frac{\hbar \kappa \alpha \omega^2}
{(\alpha^2+\omega^2)} \label{B0903} \\
\int_{0}^{\infty} d \tau \cos (\omega \tau) D(\tau)
&=& - \frac{\hbar \kappa \alpha^3}{2(\alpha^2+\omega^2)} \label{B0904} \\
\int_{0}^{\infty} d \tau \sin (\omega \tau) D(\tau)
&=& - \frac{\hbar \kappa \alpha^2 \omega}{2(\alpha^2+\omega^2)}\,.
\label{B0905}
\end{eqnarray}
It can be noticed that the restrictions on $\beta \alpha$
and $\beta$ in (\ref{B0900})-(\ref{B0903}) can be roughly replaced
by conditions on $\beta \omega$.
\fi

%%%%%%%%%%%%%%%%%%%%%%%%%%%%%%%%%%%%%%%%%%%%%%%%%%%%%%%%%%%%%%%%%%%%%%%%%%%%%%%%%%%%%%%%%%%%%%%%%%%%%
\section{Ullersma's Hamiltonian \label{oscillatorHR}}

Ullersma \cite{Ullersma} and other authors \cite{Haake} work with
a modified Hamiltonian of the oscillator model,
\begin{eqnarray}
\hat{H} &=& \frac{1}{2} (\hat{P}^2 +\omega_0^2 \hat{Q}^2) +
\frac{1}{2} \!\sum_{n=1}^{N}\! (\hat{P}^2_n +\omega_n^2 \hat{Q}^2_n)
+ \hat{Q} \!\sum_{n=1}^{N} \!\epsilon_n \hat{Q}_n\,.
\nonumber%\label{E0100}
\end{eqnarray}
\iffalse
We can rewrite the total Hamiltonian as
\begin{eqnarray}
\hat{H} = \frac{1}{2} (\hat{P}^2 + \sum_{n=1}^{N} \hat{P}^2_n)
+ V(\hat{Q},\{ \hat{Q}_n \}),\nonumber
\end{eqnarray}
where
\fi
The potential-energy part
\begin{eqnarray}
V(\hat{Q},\{ \hat{Q}_n \}) &=& \frac{1}{2} \Big(\omega_0^2 \hat{Q}^2
+ \sum_{n=1}^{N} \omega_n^2 \hat{Q}^2_n\Big)
+ \hat{Q} \sum_{n=1}^{N} \epsilon_n
\hat{Q}_n \nonumber
\end{eqnarray}
has a minimum of the potential created by the other harmonic
oscillators on the central oscillator given by
\begin{eqnarray}
\frac{\partial V(\hat{Q}_0,\{ \hat{Q}_n \})}{\partial \hat{Q}_n}
\vert_{\hat{Q}_n=\hat{Q}_n({\rm min})}
= \omega_n^2 \hat{Q}_n({\rm min}) + \epsilon_n \hat{Q}_0 = 0\, .\nonumber
\end{eqnarray}
The central oscillator thus ``feels'' the potential
\begin{eqnarray}
V(\hat{Q}_0,\{ \hat{Q}_n({\rm min}) \}) &=& \Big(\frac{\omega_0^2}{2} -
\sum_{n=1}^{N} \frac{\epsilon_n^2}{2 \omega_n^2}\Big) \hat{Q}_0^2\,.\nonumber
\end{eqnarray}
Clearly, then, positivity is not manifest; rather, in order to have
bound states, we have to impose the condition
\begin{eqnarray}
\omega_0^2-\sum_{n=1}^{N}\frac{\epsilon_n^2}{\omega_n^2}
=\omega_0^2-\kappa\alpha \geq 0\,.
\label{E0200}
\end{eqnarray}

Ullersma's Hamiltonian can be mapped onto the QBM Hamiltonian by
renormalizing the frequency $\omega_0$ as
$\omega_0^2 \to \omega_0^2 + \sum_{n=1}^{N} \frac{\epsilon^2}{\omega_n^2}
=\omega_0^2+\kappa\alpha$.
%\label{E0400}\,.
%\end{eqnarray}
That mapping was extensively used above in transcribing the
rigorous results of Ref.~\cite{Haake} to the dynamics generated by the
QBR Hamiltonian.

Needless to say, we could have based our study of the
transition from normal damping to overdamping on Ullersma's model.
Only one subtlety about that alternative treatment is worth being
mentioned here. To leading order in $\omega_0/\alpha$ the critical
value $\kappa_c$ of the coupling turns out to coincide with the border
$\kappa_{\rm max}=\omega_0^2/\alpha$ to positivity loss following from
(\ref{E0200}).

%%%%%%%%%%%%%%%%%%%%%%%%%%%%%%%%%%%%%%%%%%%%%%%%%%%%%%%%%%%%%%%%%%%%%%%%%%%%%%%%%%%%%%%%%%%%%%%%%%%%%


\begin{thebibliography} {1}
\bibitem{Rem1}
To simplify the the looks of what follows we somewhat frivolusly set
the masses of all oscillators equal to unity.
\end{thebibliography}
\end{document}